1   **Corticosterone selectively decreases humoral immunity in female eiders**

2   **during incubation**

5   Sophie Bourgeon* and Thierry Raclot

9   *Institut Pluridisciplinaire Hubert Curien (IPHC), Département Ecologie, Physiologie et*

10  *Ethologie (DEPE), UMR 7178 CNRS-ULP,*

11   *23 rue Becquerel, F-67087 Strasbourg Cedex 2, France*

17  Running title: Corticosterone and immunity in birds

19  *To whom correspondence should be addressed:

20  Phone: (33) 3.88.10.69.15

21  Fax: (33) 3.88.10.69.06

22  E-mail: sophie.bourgeon@c-strasbourg.fr

1  **SUMMARY**

2  Immunity is hypothesized to share limited resources with other physiological functions and
3  this may partly account for the fitness costs of reproduction. Previous studies have shown that
4  the acquired immunity of female common eider ducks (*Somateria mollissima*) is suppressed
5  during their incubation during which they entirely fast. Corticosterone was proposed to be an
6  underlying physiological mechanism for such immunosuppression. Therefore, the current
7  study aimed to assess the effects of exogenous corticosterone on acquired immunity in captive
8  eiders. To this end, females were implanted with corticosterone pellets at different stages of
9  their incubation fast. We measured total immunoglobulin levels, T-cell-mediated immune
10  response, body mass and corticosterone levels in these females and compared them with those
11  of control females prior to and after manipulation (i.e. corticosterone pellet implantation). To
12  mimic corticosterone effects on body mass, we experimentally extended fasting duration in a
13  group of females termed 'late fasters'. Implanted females had corticosterone levels 6 times
14  higher and lost 35 % more mass than control females. Corticosterone levels in 'late fasters'
15  were similar to those in control females but body mass was 8 % lower in the former. The
16  decrease in the immunoglobulin levels of corticosterone implanted females was twice as high
17  as in control females, while the T-cell-mediated immune response was not significantly
18  affected by the treatment. We found a decrease in the T-cell-mediated immune response only
19  in 'late fasters' (by 60 %), while the immunoglobulin level was not lower in this group than in
20  corticosterone implanted or control females. Our study shows that in incubating eiders,
21  exogenous corticosterone decreased only humoral immunity. We suggest that the
22  immunosuppressive effect of corticosterone could be mediated through its effects on body
23  reserves. Further experiments are required to determine the relationship between body
24  condition and immune system in fasting birds.

25  **Key-words:** Birds; Body reserves; Fasting; Glucocorticoids; Immunosuppression

**INTRODUCTION**

Life-history theory assumes that a major trade-off occurs between reproduction and survival, so that the costs associated with a given reproductive effort might have a deleterious impact on adult survival and future reproduction (Williams, 1966; Stearns, 1992). This concept is based on physiological trade-offs between resource-demanding activities within an individual. Recently, the costs of immune defenses have been emphasized (Gustafsson et al., 1994; Sheldon and Verhulst, 1996; Råberg et al., 1998). Namely, parental effort may induce immunosuppression in birds (Moreno et al. 1999). Acquired immunity can be classified into humoral immunity (mediated by B-lymphocytes) and cell-mediated immunity (mediated by T-lymphocytes) (Roitt et al., 1998). Deerenberg et al. (1997) showed in captive zebra finches (*Taeniopygia guttata*) that their humoral immunity was progressively reduced when the reproductive effort is increased. Similarly, when raising experimentally enlarged broods, female pied flycatchers (*Ficedula hypoleuca*) exhibited reduced T-cell-mediated responses (Moreno et al., 1999). Previous studies have shown that the acquired immunity of female common eiders (*Somateria mollissima*) is suppressed during the incubation fast (Hanssen et al., 2004; Bourgeon et al., 2006), while its experimental activation has strong negative effects on the fitness of female eiders (Hanssen et al., 2004). Whatever the ultimate factors explaining such an immunosuppression in breeding eider ducks, the underlying physiological mechanisms still remain unknown. Among potential proximate factors underlying immunosuppression during reproduction, a link between immunocompetence and hormonal changes has been proposed (see Deerenberg et al., 1997).

Glucocorticoids are an essential component of the endogenous immunoregulatory network, while also being associated with stress. Hence, these hormones establish a close endocrine link between immunocompetence and stress (Apanius, 1998). Råberg et al. (1998)

hypothesized that corticosterone, secreted during stressful activities, reduces the acquired immune function. However, while the T-cell-mediated immune response was suppressed by experimentally elevated corticosterone levels in non-breeding New-Jersey house sparrows (*Passer domesticus*; Martin II et al., 2005), it did not significantly covary with natural corticosterone concentrations in breeding barn swallows (*Hirundo rustica*; Saino et al., 2002). Similarly, Bourgeon et al. (2006) found that in female common eiders both components of the acquired immunity decreased independently of plasma corticosterone levels, which itself did not vary significantly over the incubation period. However, in several bird species, organisms are metabolically prepared for a long-term fast (Le Maho et al., 1981; Cherel et al., 1988; Lindgård et al., 1992). Indeed, fasting is first characterized by glycogen reserves exhaustion (phase I) before a long period of protein sparing and preferential mobilization of fat stores (phase II), which is followed by a period of increased net protein catabolism (phase III). While phases I/II are characterized by the maintenance of low plasma levels of corticosterone, phase III of fasting is associated with an increase in plasma corticosterone levels. Such an elevation could be responsible for the increase in protein catabolism (proteolysis), because corticosterone is known to mobilize peripheral calorie stores for glucose production and energy utilization (Robin et al., 1998). Depending on conditions during the breeding season, incubating female eiders can enter phase III of fasting (see Le Maho et al., 1981; 1983).

Our main objective in this study was to examine the effects of increased plasma corticosterone levels on both components of the acquired immunity in wild female common eiders at different stages of their incubation fast. To this end, female eiders, nesting in the high arctic, were implanted with corticosterone pellets both at the beginning and at the end of incubation. Subsequently, female total immunoglobulin levels, T-cell-mediated immune response, body mass, and plasma corticosterone were measured and compared with those in control birds prior to and after manipulation. To mimic corticosterone effects on body mass, fasting duration in a group of females (termed 'late fasters') was experimentally extended. Body mass loss in these females was therefore increased, while corticosterone levels remained unaltered (birds remained in phase II and did not reach phase III). This allowed us to discriminate the effects of increased body mass loss on the acquired immunity from those of elevated corticosterone levels. We predicted that an increased plasma corticosterone level, which increases proteolysis during the female fast, should have immunosuppressive effects. Implanted females should therefore show T-cell-mediated immune responses and/or an immunoglobulin level lower than in control females but similar to that of 'late fasters'.

## MATERIALS AND METHODS

The study was conducted in a common eider colony on Prins Heinrich Island, Kongsfjorden, Svalbard Archipelago (78°55' N, 20°07' E) between June and July 2005. The breeding colony in the study site contained about 105 nests. Females in the study area laid between one and five eggs, but a clutch size of three to four eggs was most common (20 % and 61 %, respectively, N=105). Eider ducklings are precocial and are cared for by the female only. Incubation lasts between 24 and 26 days (Korschgen, 1977). All birds started their incubation between June 07 and June 15, the main laying period for the colony. Ducks that laid their eggs after this period were not considered in this study. Ambient temperatures in June and July ranged from 2 to 10°C.

Sampling protocol

Nests were checked at least every second day throughout the study period. This was done to determine initial clutch size but also to investigate the rate of egg predation and nest desertion. A clutch of eggs was considered complete when no additional egg was laid during a two-day period (Yoccoz et al., 2002). Females which suffered partial egg predation were excluded from this study. Female eiders were caught on their nests using a bamboo pole with a nylon snare. Blood was collected from the brachial vein within three minutes of capture. It was stored in tubes containing EDTA (an anticoagulant agent) and kept on ice until being centrifuged in the laboratory (at 10, 000 rpm, for five minutes, at 4°C). Plasma samples were stored at –20°C and subsequently used to measure immunoglobulin and corticosterone levels. After blood sampling, body size was recorded (wing and tarsus lengths) and birds were weighed with a portable electronic balance (± 2 g).

Experimental groups and corticosterone implantation

A total of 36 females with mean clutch sizes of three to four eggs (3.69 ± 0.08 eggs, N=36) (mean ± SE) was used in this study. Freely incubating females were classified into two experimental groups: corticosterone (N=18) and control females (N=18). Because previous results have shown that both components of the acquired immune system are decreased during incubation in eider ducks (Bourgeon et al., 2006), females from both groups were caught during the first part of incubation (10.64 ± 0.62 days, N=22) (mean ± SE) (11 females from each group) and near the end of incubation (20.07 ± 0.24 days, N=14) (7 females from each group). After capture, a blood sample was taken, and body size and mass were recorded. Half of the females were then implanted with corticosterone pellets (see below), while the others underwent the same procedure without actual implantation. Birds were held in cages for the following 5 days at ambient temperatures, with snow given as fresh water *ad libitum*. Four days after the treatment (18.31 ± 0.87 days into their incubation, N=36) (mean ± SE), another blood sample was taken, a PHA skin test conducted (see below), and body mass determined. Birds were released 24 hours later, after the PHA skin test had been read. Additionally, in a group termed 'late fasters', which consisted of 7 captive females with mean clutch sizes from two to five eggs (3.43 ± 0.53 eggs, N=7) (mean ± SE), fasting duration was experimentally extended. Birds remained in phase II of fasting during experimentation and never entered into phase III. This protocol allowed to discriminate the effects of increased body mass loss on the acquired immunity from those of elevated corticosterone levels. To this end, these females had been incubating eggs for at least 23 days (24.86 ± 0.40 days, N=7) (mean ± SE), i.e. eggs were close to hatching. We took a blood sample and recorded body size and mass. Females were then held in captivity for five days (29.86 ± 0.40 days into their incubation, N=7) (mean ± SE), at which point a further blood sample was taken, a PHA skin test performed (see below), and body mass recorded. Females were released 24 hours after this, when the PHA skin test had been read.

Corticosterone pellets (100 mg, 21 day release, G-111) were obtained from Innovative Research of America (Sarasota, Florida, USA). The implants slowly release the hormone which enters the bloodstream. In preliminary trials we found that this dose was sufficient to induce marked increases in the level of plasma corticosterone and accelerate body mass loss as early as two days after the start of treatment (S. Bourgeon et al., unpublished data). For the implantation of the corticosterone pellets we followed the recommendations of the manufacturer. Briefly, a small patch of skin at the back side of the birds' neck was shaved and disinfected using alcohol and betadine (iodine solution). A small incision equal to the size of the pellet was made and the implant was inserted underneath the skin. The skin was closed with a single stitch, using surgical thread. The wound was cleaned with betadine and sprayed with an aluminium powder. The surgical procedure required less than 10 minutes. Control animals underwent the identical procedure, without actual insertion of a pellet.

1  T-cell-mediated immune response: PHA skin test

2  For this test we challenged one wing-web with the mitogenic phytohemagglutinin (PHA),
3  while the other wing-web (control) was injected with phosphate buffered saline (PBS).
4  Briefly, 100 µl of 5 mg.ml$^{-1}$ PHA (Sigma L 8754) in PBS were injected intradermally into the
5  right wing-web, while the left wing-web was injected with an equal volume of PBS. This
6  procedure was shown to induce little physiological stress in birds (Merino et al., 1999).
7  Injection sites on the wing-web were measured with a micrometer calliper (three readings)
8  just before and 24 h after injection with PHA or PBS. The T-cell-mediated immune response
9  was taken as the difference between the two wing-web swellings.

11  Immunoglobulin levels: ELISA test

12  A sensitive ELISA method was used to determine the amount of serum immunoglobulins in
13  eider duck blood. This method using commercial anti-chicken antibodies has so far been
14  validated in six wild avian species (Martinez et al., 2003). Despite the fact that Anseriforms
15  have an additional immunoglobulin isotype (IgY), which is not found in other birds (Parham,
16  1995), we assumed linear cross-reactivity. Accordingly, the values obtained were used as
17  relative immunoglobulin levels.
18  To determine the linear range of the sigmoid curve, ELISA plates were coated with serial
19  dilutions of serum (100 µl) in carbonate–bicarbonate buffer (0.1 M, pH = 9.6) and incubated
20  overnight at 4°C. We selected the data obtained from trials using the serum dilution nearest to
21  the centre of its linear range. ELISA plates were then coated with 100 µl of diluted serum
22  samples from female eiders (two samples per female diluted to 1/32000 in carbonate-
23  bicarbonate buffer) and incubated for 1 h at 37°C. After a second incubation overnight at 4°C,
24  the plates were washed with a solution (200 µl) of phosphate buffer saline and Tween (PBS-
25  Tween), and a diluent (100 µl), containing 5% powdered milk in PBS was added. Following incubation for 1 h at 37°C, the plates were washed with PBS-Tween buffer. 100 µl of anti-chicken conjugate (Sigma A 9046) was added at 1:250 and the plates were incubated for 2 h at 37°C. After three washes, the plates were filled with 100 µl of a solution consisting of 2,2'-azino-bis-(3-ethylbenzthiazoline-6-sulphonic acid) (ABTS) and concentrated hydrogen peroxide diluted to 1:1000. Following incubation for 1 h at 37°C, absorbance was measured at 405 nm using a plate spectrophotometer (Awareness Technology, Inc., Palm City, FL 34991, USA).

Assessment of the corticosterone levels

Corticosterone concentrations were determined by radioimmunoassay (RIA) in our laboratory using an $^{125}$I RIA double antibody kit from ICN Biomedicals (Costa Mesa, CA, USA). The corticosterone RIA had an intra-assay variability of 7.1 % (N=10 duplicates) and an inter-assay variability of 6.5 % (N=15 duplicates).

Statistical analysis

Statistical analysis was conducted with SPSS 12.0.1 (SPSS Inc., Chicago, IL, USA). Values are means ± standard error (SE). Corticosterone levels were not normally distributed (Kolmogorov-Smirnov test, P<0.05). Hence, data were log transformed to meet parametric assumptions, before parametric tests were used. Repeated measures two-way ANOVA was used to test for the effects of the treatment and incubation stage on corticosterone levels, body mass and immunoglobulin level. Two-way ANOVA was used to test for the effects of the treatment and incubation stage on the T-cell-mediated immune response. Linear regression analysis was used to assess the relationships between all parameters measured.

**RESULTS**

Tables 1 and 2 provide biological, hormonal and immune profiles for the female eider ducks (before and after implantation) sampled during the first part and near the end of their incubation period, respectively. Initial clutch size, tarsus and wing lengths, days of incubation, body mass, immunoglobulin level and corticosterone levels were not significantly different between the experimental and control group before manipulation (i.e. corticosterone pellet implantation) at both incubation stages.

Effects of corticosterone implants on body mass:

As expected, implants induced a significant increase in corticosterone levels (repeated measures two-way ANOVA: effects of repetition: $F_{1,32}=65.95$, $P<0.0001$; effects of treatment: $F_{1,32}=54.16$, $P<0.0001$; effects of incubation stage: $F_{1,32}=0.06$, $P=0.81$; interaction: $F_{1,32}=0.73$, $P=0.40$). Four days after implantation, corticosterone levels in implanted birds were 6 times higher than in control females, independent of incubation stage. Moreover, corticosterone levels in 'late fasters' were far below the levels of implanted females. In fact, they were not significantly different from the levels in control females sampled at the end of their incubation period (T-test: t=-0.03; N=25; P=0.97). Body mass in corticosterone implanted females was significantly decreased by 18 %, while it only decreased by 12 % in control females (repeated measures two-way ANOVA: effects of repetition: $F_{1,32}=1191.35$, $P<0.0001$; effects of treatment: $F_{1,32}=30.46$, $P<0.0001$; effects of incubation stage: $F_{1,32}=0.10$, $P=0.33$; interaction: $F_{1,32}=0.78$, $P=0.38$). Four days after implantation, implanted females were significantly lighter than control females (by 8 %), independent of the incubation stage. In fact, body mass loss per day in implanted females was 35 % higher than in control females, regardless of the incubation stage (two-way ANOVA: effects of treatment: $F_{1,32}=25.89$, $P<0.0001$; effects of incubation stage: $F_{1,32}=1.29$, P=0.26; interaction: $F_{1,32}=0.01$, P=0.92). Corticosterone levels in birds during early incubation were negatively related to body mass only after the implantation (Table 3), when high corticosterone levels were associated with low body mass values. However, there was no significant relationship between these parameters in birds near the end of their incubation period, neither before nor after implantation (Table 4). Finally, body mass of 'late fasters' was not significantly different from that of implanted females but was significantly lower than in control females sampled at the end of their incubation (8 %; Table 2) (repeated measures one-way ANOVA: effects of repetition: $F_{1,18}=707.63$, P<0.0001; effects of treatment: $F_{2,18}=10.89$, P=0.001).

Effects of corticosterone on the T-cell-mediated immune response and immunoglobulin level: Corticosterone implants had no significant effect on the T-cell-mediated immune response (two-way ANOVA: effects of treatment: $F_{1,32}=2.49$, P=0.12; effects of incubation stage: $F_{1,32}=2.94$, P=0.10; interaction: $F_{1,32}=0.003$, P=0.95; Fig. 1). Responses in implanted females were similar to that of control females, independent of the incubation stage. However, the immune response in 'late fasters' was significantly reduced when compared with implanted and control females at the end of their incubation (53 % and 63 %, respectively) (one-way ANOVA: effects of treatment: $F_{2,18}=9.39$, P=0.002; Fig. 1). There was no relationship between the T-cell-mediated immune response and plasma corticosterone levels at any stage (Tables 3 and 4). However, there was a positive significant relationship between the T-cell-mediated immune response and body mass but only during early incubation (Tables 3 and 4), so that the immune response was stronger in heavier females.

After four days, the immunoglobulin level in implanted females was significantly decreased by 45 and 33 %, when sampled during early and late incubation, respectively. The corresponding decline in immunoglobulin level of control females was only 25 % and 25 % (repeated measures two-way ANOVA: effects of repetition: $F_{1,32}=105.10$, $P<0.0001$; effects of treatment: $F_{1,32}=7.34$, $P=0.01$; effects of incubation stage: $F_{1,32}=7.14$, $P=0.01$; interaction: $F_{1,32}=2.77$, $P=0.11$; Fig. 2), indicating that corticosterone pellet implantation had a negative effect on the female immunoglobulin level. This effect was stronger when implantation occurred during early incubation rather than during late incubation. The immunoglobulin level in 'late fasters' was not significantly different from that of implanted or control females sampled near the end of their incubation (repeated measures one-way ANOVA: effects of repetition: $F_{1,18}=78.15$, $P<0.0001$; effects of treatment: $F_{2,18}=0.74$, $P=0.49$; Fig. 2). We found no relationship between the immunoglobulin and plasma corticosterone levels, neither before nor after the treatment during early incubation. The same held true for the period after implantation, when birds were near the end of incubation. By contrast, both parameters were negatively related before implantation of ducks that were near the end of the incubation period. High corticosterone levels were then associated with a low immunoglobulin level (Table 4; Fig. 3). Furthermore, there was no relationship between the immunoglobulin level and body mass, neither before nor after the treatment at any stage of incubation (Tables 3 and 4). Finally, we did not find a relationship between both components of the acquired immunity after hormone implantation at any stage of the eider duck incubation period (Tables 3 and 4). Hence, the immunoglobulin level was independent of the T-cell-mediated immune response.

**DISCUSSION**

It has previously been shown that the acquired immunity is significantly decreased during the incubation fast of female common eiders (Hanssen et al., 2004; Bourgeon et al., 2006). Among other potential scenarios, reduced immunocompetence during reproduction has been attributed to hormonal regulation, in particular through the action of glucocorticoids (Deerenberg et al., 1997; Råberg et al., 1998). Hence, the main objective of the current study was to investigate the potential physiological mechanisms underlying such immunosuppression. We therefore assessed the effects of exogenous corticosterone on both components of the acquired immune system in female eiders during different stages of their incubation fast. In addition, a group of females whose fasting duration was experimentally extended was used to discriminate the effects of increased body mass loss on the acquired immunity from those of elevated corticosterone levels.

Experimentally increased plasma corticosterone levels only affected one of the two components of the acquired immunity. The immunoglobulin level in implanted females was significantly decreased when compared with that of control females. This response was strongest when birds were implanted at the beginning of their incubation fast. However, the T-cell-mediated immune response was not significantly affected by the treatment. Paradoxically, there was no significant relationship between plasma corticosterone and immunoglobulin levels after the implantation at any incubation stage. By contrast, before the treatment in ducks that were near the end of their incubation, high corticosterone levels seemed to be associated with a low immunoglobulin level.

Whatever the effects of exogenous corticosterone and similar to an earlier investigation (Bourgeon et al., 2006), we did not find a significant relationship between both components of the acquired immunity. This lends support to the view that variations in one component of the acquired immunity are not necessarily a reliable indicator of changes in the other (Norris and Evans, 2000). In fact, in the present study, the immunoglobulin level seemed to be more sensitive to corticosterone treatment than the T-cell-mediated immune response. In control females, the immunoglobulin level significantly decreased throughout incubation, while the T-cell-mediated immune response did not vary significantly. This would seem to contrast with the finding of a previous investigation (Bourgeon et al., 2006) supporting that both components significantly decrease throughout the incubation fast of eider ducks. This apparent discrepancy could hold in the fact that smaller sample sizes have been used, what is reinforced by high variances observed in this immune response. Moreover, we can not exclude the possibility that effects of corticosterone on the T-cell-mediated immune response might require more time (see Dhabhar and Mc Ewen, 1997) and/or higher corticosterone concentrations. In the current study, a significant decrease in the T-cell-mediated immune response was only observed in 'late fasters', while their immunoglobulin level was not lower than that of corticosterone implanted or control females. Implanted females significantly lost more weight than control females, which is consistent with the observation by Cherel et al. (1988) that high levels of corticosterone increase proteolysis in fasting birds. In a preliminary study on captive female eiders, high doses of exogenous corticosterone, administered for a few days, induced a rise in plasma levels of uric acid, indicating protein breakdown (Criscuolo et al., 2005). Corticosterone levels of 'late fasters' in the current study were similar to that of control females but body mass was 8 % lower in the former. Hence, the T-cell-mediated immune response appears to be more sensitive to body mass loss than to elevated levels of corticosterone, supporting the view of an indirect effect of corticosterone on this immune parameter. Accordingly, in the present study we found a positive relationship between body mass and T-cell-mediated immune response, where the response was stronger in females with a greater body mass. By contrast, the immunoglobulin level appears to be more sensitive to high corticosterone levels than to body mass loss. Nevertheless, corticosterone levels and body mass were negatively related in the current study, so that high levels of corticosterone were associated with low body mass. Our results agree with previous findings from breeding black-legged kittiwakes (*Rissa tridactyla*), where high corticosterone levels were associated with a marked decline in body condition (Kitaysky et al., 1999). In this context, it would be interesting for future studies 1) to experimentally extend the fasting duration of eiders until they reach phase III of fasting or 2) to find free-ranging birds which spontaneously shift from lipid to protein utilization, so that the effects of both elevated corticosterone levels and decreased body mass on the acquired immunity can be examined.

The present study showed that an increasing body mass loss, caused either by corticosterone administration, or by an experimental extension of fasting duration, negatively affected the birds' immunoglobulin level and their T-cell-mediated immune response, respectively. This result lends support to the resource-limitation hypothesis which predicts that investment in costly behaviours, such as reproduction, reduces the amount of resources available to other systems, such as the immune system (Sheldon and Verhulst, 1996; Råberg et al., 1998). However, evidence for an energetically costly immune response is still equivocal (Råberg et al., 1998; Eraud et al., 2005; Verhulst et al., 2005). Whatever the ultimate factors explaining immunosuppression during reproduction, corticosterone was proposed to regulate immunosuppression in incubating birds (Deerenberg et al., 1997; Råberg et al., 1998; Saino et al., 2003). However, during fasting corticosterone should be maintained at low levels to avoid metabolism disorders, such as protein catabolism (Cherel et al., 1988). In the current study we did not find marked variations in corticosterone levels during the incubation period of control and 'late fasting' females. However, 'late fasters' had a lower T-cell-mediated immune responses than corticosterone implanted or control females. This would support the view that fasting duration and/or body composition (see below) might be relevant parameters for immunosuppression.

Other aspects of immunosuppression, namely humoral immunity, could be mediated by factors related to fuel utilization or body mass loss. Such a relationship between energy storage/mobilization and immunocompetence might plausibly be mediated through nutritional and/or endocrine factors (Apanius, 1998). Exogenous administration of corticosterone is likely to increase proteolysis. Consequently, lean body mass will be decreased, while the energy stored as lipids within the body will be spared. Hence, for the same final body mass, adiposity of 'late fasters' should be lower than for corticosterone implanted females as reported to be the case in dark-eyed juncos (*Junco hyemalis*) treated with corticosterone (Gray et al., 1990). Currently, adipose tissue is perceived as an active participant in the regulation of essential and prominent body processes such as immune homeostasis (Matarese and La Cava, 2004). This raises the question of how body reserves might control the immune system (Demas and Sakaria, 2005). Some adipose humoral signals, such as leptin, are generated in proportion to fat stores and act on feedback control systems to influence numerous biological processes (Lõhmus and Sundtröm, 2004; Matarese et al., 2005). In fact, leptin is secreted primarily by adipose tissue and has been shown to enhance a variety of immunological parameters in mammals (Lord et al., 1998; Faggioni et al., 2001; Demas et al., 2003) and birds (Lõhmus et al., 2004). Consequently, leptin levels might be lower in 'late fasters' when compared with corticosterone implanted eiders. To gain further insight into the role that leptin plays for the immune system of fasting birds, plasma measurements of leptin and manipulation of its circulating concentrations would be useful.

In conclusion, exogenous corticosterone decreased only one component of the acquired immune system in incubating female eider ducks. While the treatment significantly decreased their immunoglobulin level, their T-cell-mediated immune response was not affected. Implanted females lost significantly more weight than control birds. Females, whose fasting duration was experimentally extended, increasing body mass loss, displayed lower T-cell-mediated immune responses than implanted females, while their corticosterone levels remained at baseline values. Consequently, the immunosuppressive effect of corticosterone appears to be mediated by its effect on body reserves, which have been shown to play an important role in the regulation of the immune system. For example, leptin, which conveys 1 information on energy availability, could be involved in the observed immunosuppression.
2 Further experiments are required to determine the relationship between body condition and
3 immune function in incubating female eiders. Our results raise the question of the
4 physiological mechanisms which can explain the effects of corticosterone on the immune
5 response. Furthermore, could it be that depending on its concentration, this hormone is able to
6 trigger different responses, as has already been reported in the context of foraging behaviour
7 (Wingfield et al., 1998).
8
9
10
11
12
13
14
15
16
17
18
19
20
21
22
23
24
25

**ACKNOWLEDGMENTS**

Financial support for this study was provided by the Institut polaire français Paul Emile Victor (IPEV) and through a fellowship from the French MENRT to S. Bourgeon. The study was approved by the Norwegian Animal Research Authority and the Ethic Committee of the Institut Polaire Français Paul-Emile Victor. Permission to work on Common Eiders was also granted by the Governor of Svalbard.

10
11
12
13
14
15
16
17
18
19
20
21
22
23
24
25

**Table 1.** Profiles for both experimental groups of captive female eiders (corticosterone and control group) which were sampled during their early phase of incubation. 'Time of sampling' indicates incubation stage. Values are means ± SE. Lower case a and b indicate a significant difference between groups (T-tests).

| Before implantation | Group 1: Corticosterone females (N=11) | Group 2: Control females (N=11) |
|---|---|---|
| Initial clutch size (eggs) | 3.82 ± 0.12[a] | 3.54 ± 0.16[a] |
| Tarsus length (cm) | 6.15 ± 0.05[a] | 6.21 ± 0.04[a] |
| Wing length (cm) | 29.14 ± 0.15[a] | 29.45 ± 0.17[a] |
| Time of sampling (days) | 10.27 ± 0.76[a] | 11.00 ± 1.00[a] |
| Body mass at sampling (g) | 1687 ± 39[a] | 1753 ± 34[a] |
| Immunoglulin level (absorbance units) | 1.04 ± 0.06[a] | 0.94 ± 0.07[a] |
| Corticosterone (ng.ml$^{-1}$) | 14.34 ± 2.17[a] | 14.11 ± 1.63[a] |
| **After implantation** | **Group 1: Corticosterone females (N=11)** | **Group 2: Control females (N=11)** |
| Time of sampling (days) | 14.27 ± 0.76[a] | 15.00 ± 1.00[a] |
| Body mass at sampling (g) | 1430 ± 33[a] | 1575 ± 32[b] |
| Total body mass loss (g) | 257.09 ± 13.12[a] | 178.55 ± 8.42[b] |
| Immunoglulin level (absorbance units) | 0.57 ± 0.05[a] | 0.70 ± 0.05[a] |
| T-cell-mediated immune response (mm) | 0.72 ± 0.14[a] | 0.97 ± 0.18[a] |
| Corticosterone (ng.ml$^{-1}$) | 109.90 ± 8.16[a] | 14.70 ± 1.85[b] |

**Table 2.** Profiles for the three experimental groups of captive female eiders (corticosterone, control, and 'late fasters'), which were sampled near the end of their incubation. 'Time of sampling' indicates incubation stage. Values are means ± SE. Lower case a, b and c indicate a significant difference between groups (one-way ANOVA).

| Before implantation | Group 1: Corticosterone females (N=7) | Group 2: Control females (N=7) | Group 3: Late fasting females (N=7) |
|---|---|---|---|
| Initial clutch size (eggs) | 3.71 ± 0.18$^a$ | 3.71 ± 0.18$^a$ | 3.43 ± 0.53$^a$ |
| Tarsus length (cm) | 6.16 ± 0.05$^a$ | 6.07 ± 0.04$^a$ | 6.01 ± 0.06$^a$ |
| Wing length (cm) | 29.19 ± 0.24$^a$ | 29.14 ± 0.20$^a$ | 29.29 ± 0.18$^a$ |
| Time of sampling (days) | 20.29 ± 0.29$^a$ | 19.86 ± 0.40$^a$ | 24.86 ± 0.40$^b$ |
| Body mass at sampling (g) | 1549 ± 27$^a$ | 1571 ± 37$^a$ | 1441 ± 36$^b$ |
| Immunoglulin level (absorbance units) | 0.70 ± 0.06$^a$ | 0.70 ± 0.04$^a$ | 0.82 ± 0.07$^a$ |
| Corticosterone (ng.ml$^{-1}$) | 18.36 ± 3.09$^a$ | 18.35 ± 4.45$^a$ | 13.26 ± 1.99$^a$ |
| **After implantation** | **Group 1: Corticosterone females (N=7)** | **Group 2: Control females (N=7)** | **Group 3: Late fasting females (N=7)** |
| Time of sampling (days) | 24.29 ± 0.29$^a$ | 23.86 ± 0.40$^a$ | 29.86 ± 0.40$^b$ |
| Body mass at sampling (g) | 1315 ± 23$^{a,b}$ | 1394 ± 36$^a$ | 1285 ± 27$^b$ |
| Total body mass loss (g) | 234.00 ± 15.93$^a$ | 177.14 ± 9.54$^b$ | 155.43 ± 10.45$^b$ |
| Immunoglulin level (absorbance units) | 0.47 ± 0.03$^a$ | 0.52 ± 0.04$^a$ | 0.57 ± 0.05$^a$ |
| T-cell-mediated immune response (mm) | 1.00 ± 0.13$^a$ | 1.27 ± 0.18$^a$ | 0.47 ± 0.08$^b$ |
| Corticosterone (ng.ml$^{-1}$) | 111.10 ± 18.06$^a$ | 22.33 ± 5.29$^b$ | 16.72 ± 2.21$^b$ |

1  **Table 3.** Results for linear regressions between immune parameters, body mass, and
2  corticosterone level in captive female eiders, sampled during the first part of incubation. Signs
3  given into brackets indicate positive or negative relationships.

| Before implantation | Body mass (g) | Immunoglulin level (absorbance units) | Corticosterone (ng.ml$^{-1}$) |
|---|---|---|---|
| **Body mass (g)** | - | $F_{1,21}$=1.19, P=0.29 (+) | $F_{1,21}$=0.07, P=0.80 (+) |
| **Immunoglulin level (absorbance units)** |  | - | $F_{1,21}$=3.99, P=0.06 (+) |
| **Corticosterone (ng.ml$^{-1}$)** |  |  | - |

| After implantation | Body mass (g) | Immunoglulin level (absorbance units) | T-cell-mediated immune response (mm) | Corticosterone (ng.ml$^{-1}$) |
|---|---|---|---|---|
| **Body mass (g)** | - | $F_{1,21}$=2.37, P=0.14 (+) | $F_{1,21}$=5.86, P=0.02 (+) | $F_{1,21}$=6.80, P=0.02 (-) |
| **Immunoglulin level (absorbance units)** |  | - | $F_{1,21}$=3.58, P=0.07 (+) | $F_{1,21}$=2.68, P=0.12 (-) |
| **T-cell-mediated immune response (mm)** |  |  | - | $F_{1,21}$=1.13, P=0.30 (-) |
| **Corticosterone (ng.ml$^{-1}$)** |  |  |  | - |

1 **Table 4.** Results for linear regressions between immune parameters, body mass, and
2 corticosterone level in captive female eiders, sampled near the end of incubation. Signs given
3 into brackets indicate positive or negative relationships.

| Before implantation | Body mass (g) | Immunoglulin level (absorbance units) | Corticosterone (ng.ml$^{-1}$) |
|---|---|---|---|
| **Body mass (g)** | - | $F_{1,20}$=3.00, P=0.10 (-) | $F_{1,20}$=0.26, P=0.61 (-) |
| **Immunoglulin level (absorbance units)** | | - | $F_{1,20}$=7.70, P=0.01 (-) |
| **Corticosterone (ng.ml$^{-1}$)** | | | - |

| After implantation | Body mass (g) | Immunoglulin level (absorbance units) | T-cell-mediated immune response (mm) | Corticosterone (ng.ml$^{-1}$) |
|---|---|---|---|---|
| **Body mass (g)** | - | $F_{1,20}$=4.16, P=0.05 (-) | $F_{1,20}$=0.17, P=0.68 (+) | $F_{1,20}$=0.43, P=0.52 (-) |
| **Immunoglulin level (absorbance units)** | | - | $F_{1,20}$=1.16, P=0.29 (-) | $F_{1,20}$=2.76, P=0.11 (-) |
| **T-cell-mediated immune response (mm)** | | | - | $F_{1,20}$=1.04, P=0.32 (+) |
| **Corticosterone (ng.ml$^{-1}$)** | | | | - |

**FIGURE LEGENDS**

**Fig 1.** Effects of treatment on wing-web swelling in female eiders: corticosterone implanted (hatched bars), sham implanted (plain bars), and 'late fasting' females (cross hatched bars). Values are means ± SE. Lower case a and b indicate a significant difference between groups (LSD post-hoc tests).

**Fig 2.** Effects of treatment on immunoglobulin level before (hatched bars) and after (plain bars) manipulation in female eiders. Values are means ± SE. Lower case a, b, c and d indicate a significant difference between groups (LSD post-hoc tests).

**Fig 3.** Relationship between corticosterone level and immunoglobulin level before implantation in female eiders sampled near the end of incubation.

1 **FIGURE 1.**

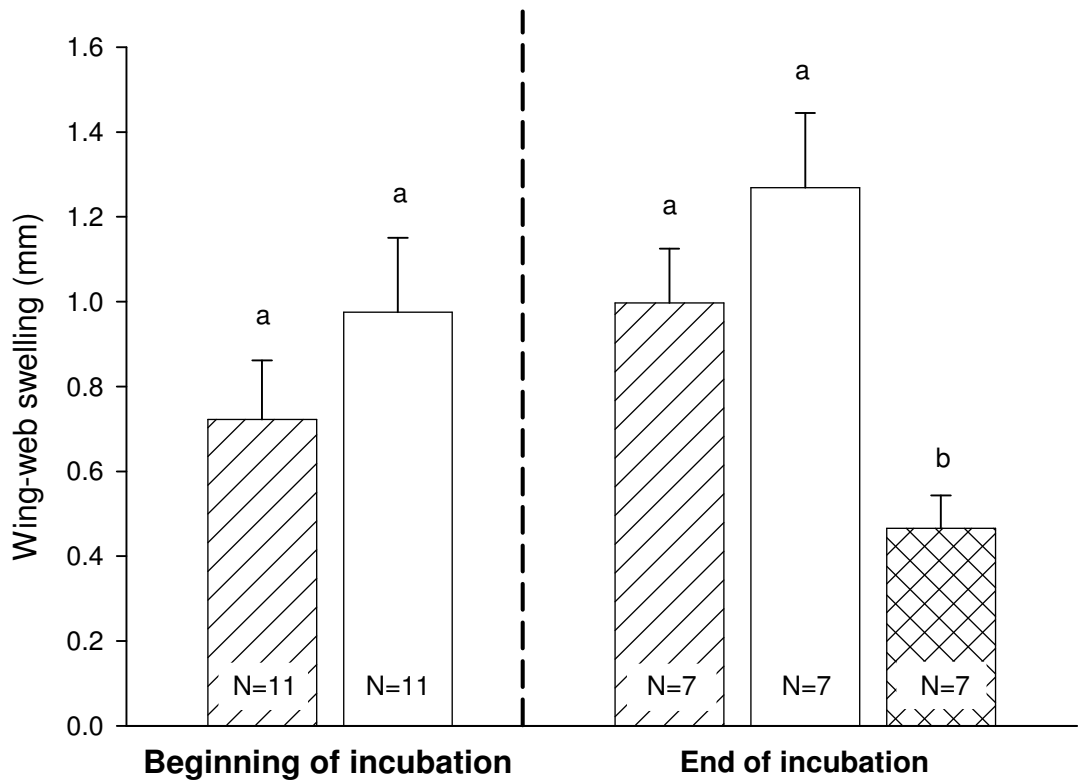

1  **FIGURE 2.**

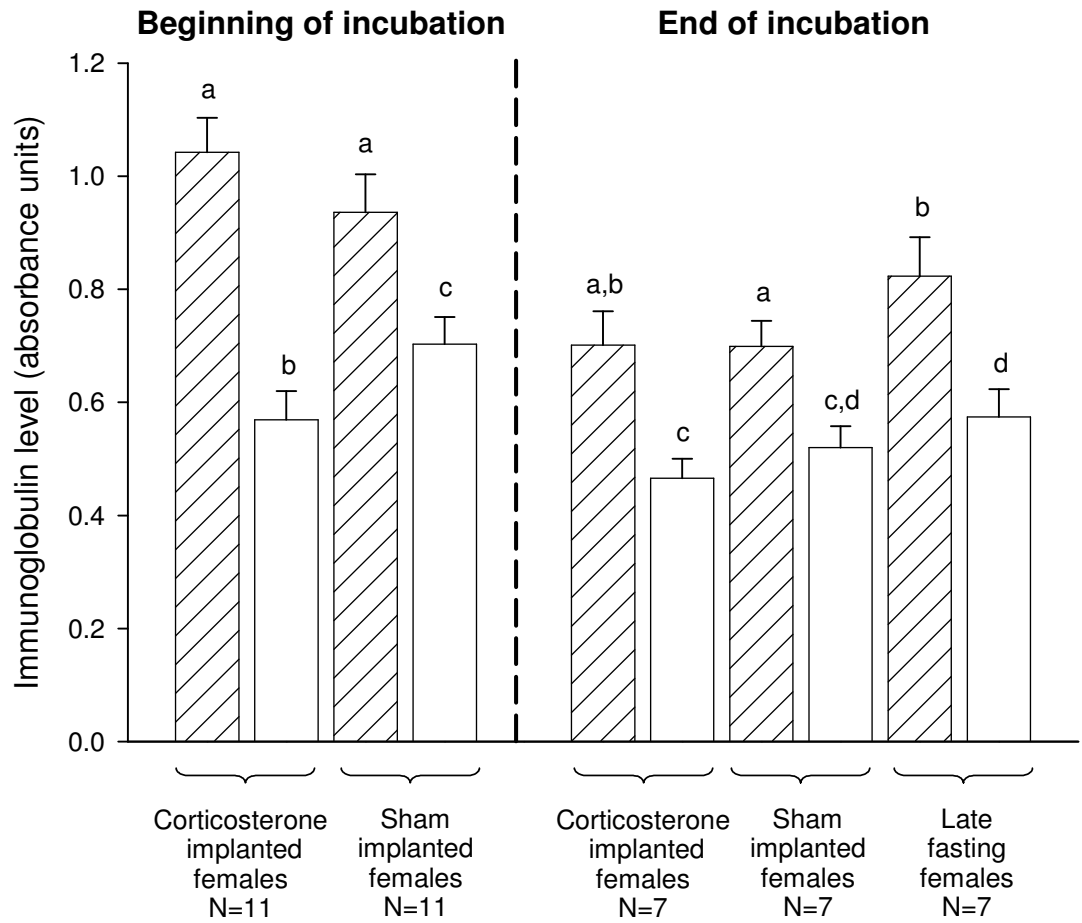

1 **FIGURE 3.**

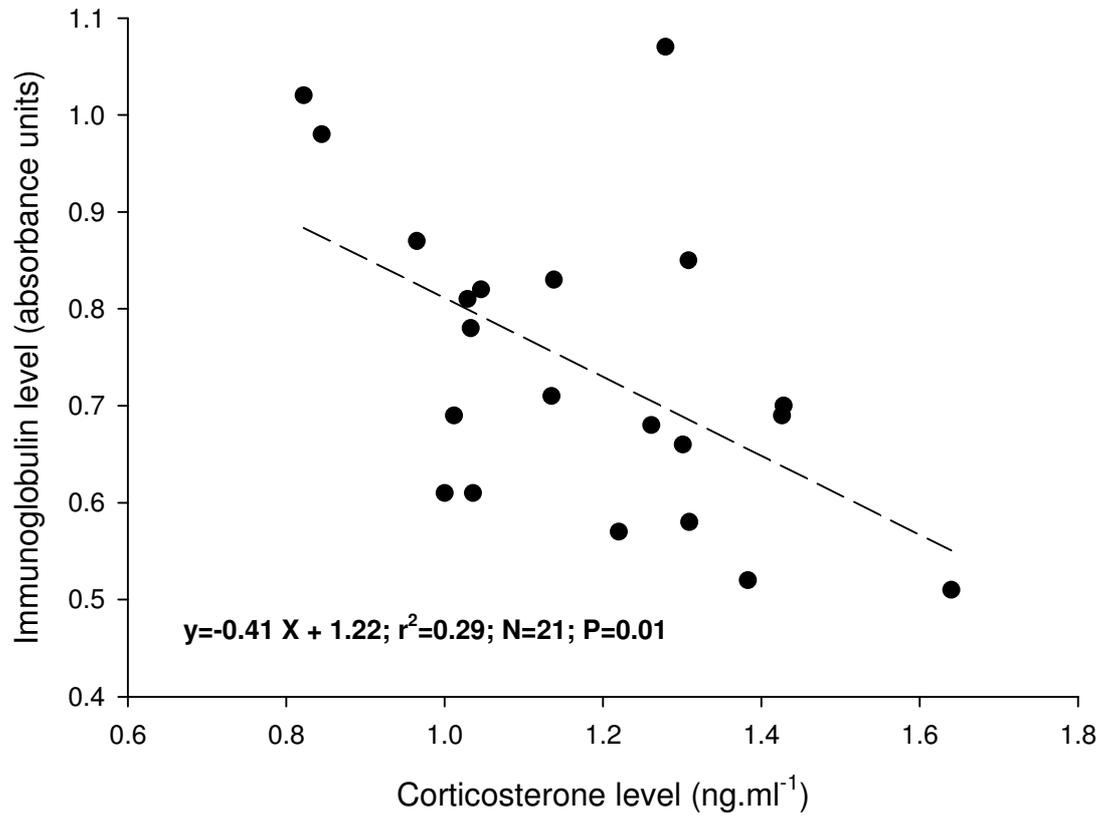

$y = -0.41 X + 1.22$; $r^2 = 0.29$; $N = 21$; $P = 0.01$

Corticosterone level (ng.ml$^{-1}$)

Immunoglobulin level (absorbance units)